%% file: main.tex
\documentclass[9pt,conference]{IEEEtran}

\usepackage[preprint]{waspaa25}

\usepackage{bm} 
\usepackage{subfig}
\usepackage{enumitem}

\title{Physics-Informed Transfer Learning for Data-Driven Sound Source Reconstruction in Near-Field Acoustic Holography}

\author{
    \IEEEauthorblockN{Xinmeng Luan}  
    \IEEEauthorblockA{
        \textit{Computational Acoustic Modeling Laboratory},\\ \textit{Center for Interdisciplinary Research in Music Media} \\ \textit{and Technology,e} \\
        Montreal, Canada \\
        xinmeng.luan@mail.mcgill.ca
    } 
    \and
    \IEEEauthorblockN{Mirco Pezzoli, Fabio Antonacci, Augusto Sarti}  
    \IEEEauthorblockA{
        \textit{Dipartimento di Elettronica, Informazione e Bioingegneria } \\
        \textit{(DEIB), Politecnico di Milano} \\
         Milan, Italy \\
        \{mirco.pezzoli, fabio.antonacci, augusto.sarti\}@polimi.it
    }
}

\name{Xinmeng Luan$^{1}$,
      Mirco Pezzoli$^{2}$,
      Fabio Antonacci$^{2}$,
      Augusto Sarti$^{2}$\thanks{}}
\address{$^{1}$Computational Acoustic Modeling Laboratory,  Center for Interdisciplinary  Research in Music Media and Technology, \\McGill University, Montreal, Canada; \\
$^{2}$Dipartimento di Elettronica, Informazione e Bioingegneria (DEIB), Politecnico di Milano, Milan, Italy
}

\begin{document}

\maketitle

\begin{abstract}

We propose a transfer learning framework for sound source reconstruction in Near-field Acoustic Holography (NAH), which adapts a well-trained data-driven model from one type of sound source to another using a physics-informed procedure. The framework comprises two stages: (1) supervised pre-training of a complex-valued convolutional neural network (CV-CNN) on a large dataset, and (2) purely physics-informed fine-tuning on a single data sample based on the Kirchhoff-Helmholtz integral. This method follows the principles of transfer learning by enabling generalization across different datasets through physics-informed adaptation. The effectiveness of the approach is validated by transferring a pre-trained model from a rectangular plate dataset to a violin top plate dataset, where it shows improved reconstruction accuracy compared to the pre-trained model and delivers performance comparable to that of Compressive-Equivalent Source Method (C-ESM). Furthermore, for successful modes, the fine-tuned model outperforms both the pre-trained model and C-ESM in accuracy.

\end{abstract}

\input{sections/introduction.tex}

\input{sections/background.tex}
\input{sections/method}
\input{sections/validation}

\input{sections/conclusion.tex}

\clearpage
\bibliographystyle{IEEEtran}
\bibliography{refs25}







\end{document}

%% file: sections/introduction.tex
\section{Introduction}
\label{sec:intro}

With the rise of deep learning in acoustics, several studies have explored its application in Near-Field Acoustic Holography (NAH) \cite{olivieri2020inter, olivieri2021eusipco, olivieri2021pinn, luancomplex, luan2025pinnsfd, wang20213d, wang2022research, wang2023cylindrical, chaitanya2023machine, lobato2024using, zhou2025reconstruction}. 
NAH is a widely used technique in acoustics for identifying and visualizing sound sources. It aims to reconstruct the vibrating surface velocity of a sound source from the pressure field measured by a microphone array at the so-called hologram plane, under the near-field assumption \cite{williams2000fourier}. 
Predicting surface velocity from hologram sound pressure, which involves inverting the Kirchhoff-Helmholtz (KH) integral, is a highly ill-conditioned process due to capturing the evanescent waves emitted by the sound source in near-field, thus it typically requires regularization \cite{williams2001regularization}.
Additionally, the problem is usually under-determined, meaning there are more surface points than measurements, leading to a non-unique solution subspace.

Previously, a U-Net-based convolutional neural network (CNN) framework was proposed to address the NAH problem. Initially, a standard data-driven CNN was employed to map the pressure field at the hologram plane to the source velocity field, using a large simulated dataset consisting of rectangular and violin top plates. The training was guided solely by a data loss on the velocity field \cite{olivieri2020inter, olivieri2021eusipco}. Subsequently, inspired by the concept of Physics-Informed Neural Networks (PINNs) \cite{raissi2019physics,koyama2025physics}, a hybrid data- and physics-driven approach was introduced by incorporating a physics-based loss term derived from the KH integral model of wave propagation. The resulting loss function was a linear combination of the source velocity data loss and a hologram pressure-based physics loss \cite{olivieri2021pinn}, and the model was referred to as KHCNN. Expanding on this approach, the framework was further developed using complex-valued neural networks (CVNNs) \cite{hirose2012complex, trabelsi2017deep}, leading to the formulation of the CV-KHCNN model \cite{luancomplex}.
Another data-driven 3D CNN-based framework for NAH was proposed in \cite{wang20213d, wang2022research, wang2023cylindrical}.
On the other hand, rather than directly employing NNs to model the sound field data, some studies concentrate on learning the indirect intermediate variables utilized in traditional NAH methods. 
For instance, the method proposed in \cite{chaitanya2023machine} utilizes NNs to estimate the coefficients of the equivalent sources for the Equivalent Source Method (ESM) \cite{koopmann1989method}. In \cite{lobato2024using}, an invertible NN \cite{ardizzone2018analyzing} was utilized together with the Helmholtz Least Squares (HELS) method \cite{wu1998reconstructing} to improve the inversion results.

This study explores the generalization capabilities of deep learning methods, with a particular emphasis on transfer learning. The main objective is to address the limitations of conventional data-driven approaches, which typically perform well on in-distribution data but struggle to generalize to out-of-distribution scenarios. To tackle this challenge, we propose a physics-informed fine-tuning strategy.
Previously, a transfer learning study based on KHCNN \cite{olivieri2021pinn} was presented in \cite{zhou2025reconstruction}, where the authors pre-trained the KHCNN on a large dataset of one type and then fine-tuned its decoder on a smaller dataset. Their validation focused on a relatively simple task: the model was pre-trained on small rectangular plates and fine-tuned on large rectangular plates, with the only differences between the two datasets being the plate dimensions and damping ratios. The results were promising, although the target scenario was relatively simple and not particularly challenging. 

 In this paper, we propose a physics-informed transfer learning framework for a data-driven NAH method.
It consists of two stages: a supervised pre-training of a CV-CNN on a large dataset, followed by physics-informed fine-tuning  on a single data sample. The data-driven pre-training avoids costly backpropagation through the KH integral, while the fine-tuning phase applies only physics-based constraints on hologram pressure, enabling adaptation without large datasets. We validate the approach by transferring a model trained on rectangular plates to violin top plates, where it demonstrates improved reconstruction accuracy over the pre-trained model and achieves competitive performance relative to the Compressive-ESM (C-ESM). Moreover, for successful modes, the fine-tuned model  outperforms both the pre-trained model and the C-ESM \cite{fernandez2017sparse}.

%% file: sections/background.tex
\begin{figure*}
\vspace{-0.8cm}
    \centering
\includegraphics[width=0.85\linewidth]{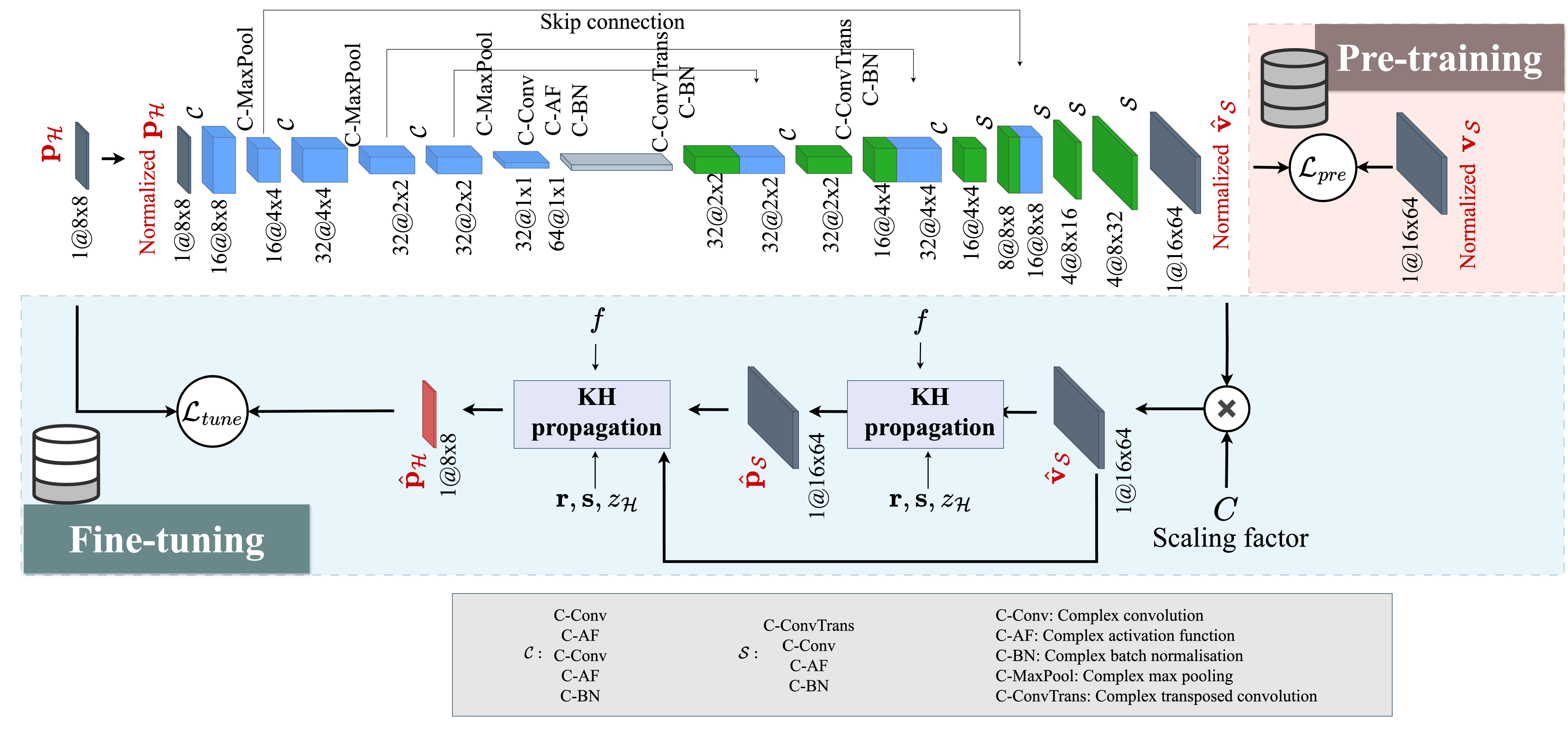}
    \caption{The framework of the proposed physics-informed transfer learning. In pre-training stage, $\textbf{p}_\mathcal{H}$ is fed in the encoder, through a bottleneck, followed by a decoder with output $\hat{\textbf{v}}_\mathcal{S}$, respectively. Then the fine-tuning stage involves the KH propagation models to get the reconstruction fields, with known frequency and configuration. }
    \label{fig:TL-PINN framework}
    \vspace{-2mm}
\end{figure*}
\section{Background on NAH}\label{sec:background}
\subsection{Kirchhoff-Helmholtz Integral}
The exterior pressure field radiated from a vibrating structure can be characterized by the well-known KH integral \cite{koopmann1989method}. 
Let $\mathcal{S}$ denote the vibrating surface with points $\mathbf{s}$, and $\mathbf{r}$ denote the exterior measured point. The pressure at $\mathbf{r}$ can be solved using the KH integral as \cite{koopmann1989method}
\begin{equation} \label{eq:kirchhoffhelmholtz}
\begin{aligned}
    p(\mathbf{r}, \omega) &= \int_\mathcal{S} p(\mathbf{s}, \omega) \frac{\partial}{\partial \mathbf{n}} g_{\omega}(\mathbf{r},\mathbf{s}) d\mathbf{s} \\ 
    &- j \omega \rho_0 \int_\mathcal{S} v_{n}(\mathbf{s},\omega) g_{\omega}(\mathbf{r},\mathbf{s}) d\mathbf{s},
\end{aligned}
\end{equation} 
where $\mathbf{n}$ is the outward normal direction unit vector, $p(\cdot,\omega)$ and $v_{\mathrm{n}}(\cdot,\omega)$ are the pressure and the normal velocity field, respectively, $\omega$ is the angular frequency and $\rho_0 \approx \SI{1.225}{\kilo\gram\per\cubic\meter}$  is the air mass density at $\SI{20}{\celsius}$.
$g_{\omega}(\mathbf{r},\mathbf{s})$ is the free-field Green's function from $\mathbf{s}$ to $\mathbf{r}$, written as \cite{williams2000fourier}
\begin{equation} \label{eq:greenfunction}
    g_{\omega}(\mathbf{r},\mathbf{s}) = \frac{1}{4\pi}\frac{e^{-j\frac{\omega}{c}\left | \left |\mathbf{r}-\mathbf{s}\right | \right |}}{\left | \left |\mathbf{r}-\mathbf{s} \right | \right |},
\end{equation}
with $c$ the sound speed in air and $j$ the imaginary unit.
\subsection{NAH Problem Formulation}
Using the KH integral, we can calculate the pressure emitted by a vibrating source based on the pressure and normal velocity fields at the surface of the object. However, NAH aims at solving the inverse problem, i.e., computing $v_n (\mathbf{s},\omega)$ starting from $p(\mathbf{r},\omega)$ acquired by a microphone array, which is a highly ill-posed problem \cite{williams2000fourier}.

In this study, we specifically focus on planar NAH. 
We define the positions $\mathbf{r}$ of $M$ measurement points located on the hologram plane $\mathcal{H}$, positioned close to the vibrating surface.
Additionally, we consider a sampled normal velocity field at positions $\mathbf{s}$, corresponding to $N$ points on the source surface plane $\mathcal{S}$.
The normal distance between plane $\mathcal{H}$ and $\mathcal{S}$ is denoted as $z_\mathcal{H}$. The inverse problem can be represented by 
\begin{equation}
    \hat{v}_n(\mathbf{s},\omega) \Big |_{\mathbf{s} \in \mathcal{S}} \approx \Gamma^{-1} \left [ p(\mathbf{r},\omega)\right] \Big | _{\mathbf{r} \in \mathcal{H}},
\end{equation}
where $\Gamma$ is a discrete estimator that approximates the pressure field on the hologram plane given the normal velocity field on the surface plane. 

%% file: sections/method.tex
\begin{figure*}
\vspace{-1cm}
    \centering
        \makebox[0pt][r]{%
        \rotatebox{90}{\hspace{1cm}\small Mode $5$, \SI{268}{\hertz}}%
    }%
    \subfloat[Ground truth]{
        \includegraphics[scale=0.28]{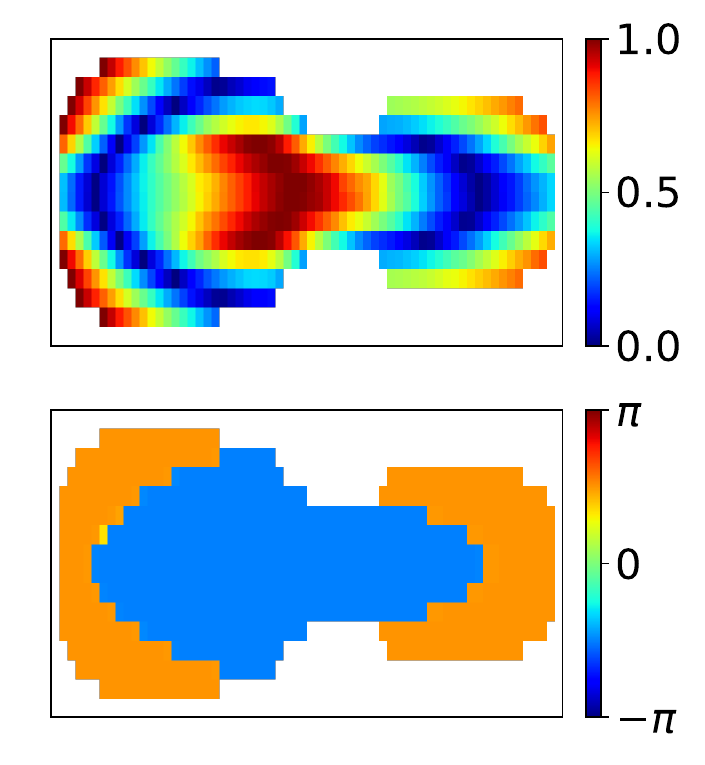}
    }
    \subfloat[\parbox{2cm}{\centering Fine-tuned \\ \tiny{NMSE: \textbf{-8.78}} \\ \tiny{NCC: \textbf{93.27\%}}}]{
        \includegraphics[scale=0.28]{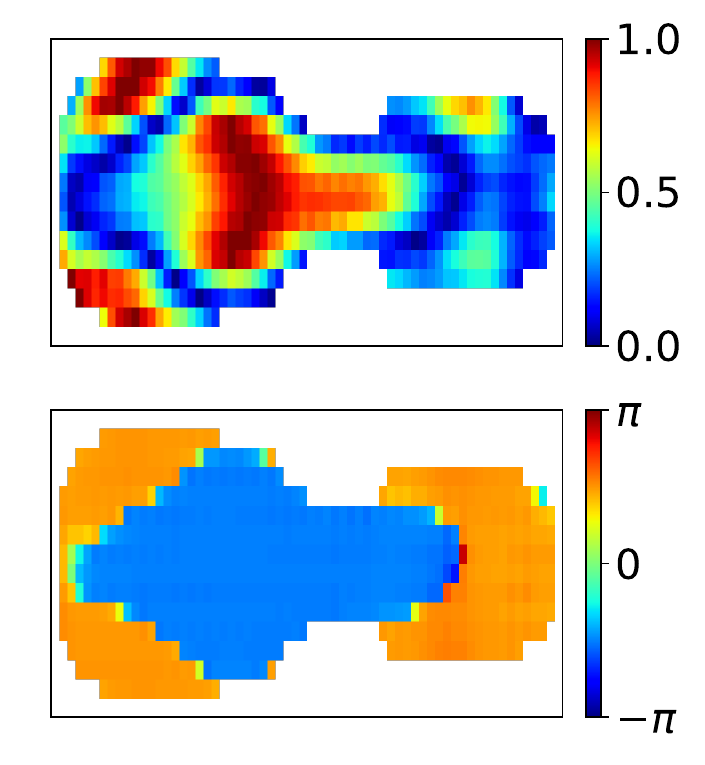}
    }
    \subfloat[\parbox{2cm}{\centering Pre-trained \\ \tiny{NMSE: -4.74} \\ \tiny{NCC: 82.70\%}}]{
        \includegraphics[scale=0.28]{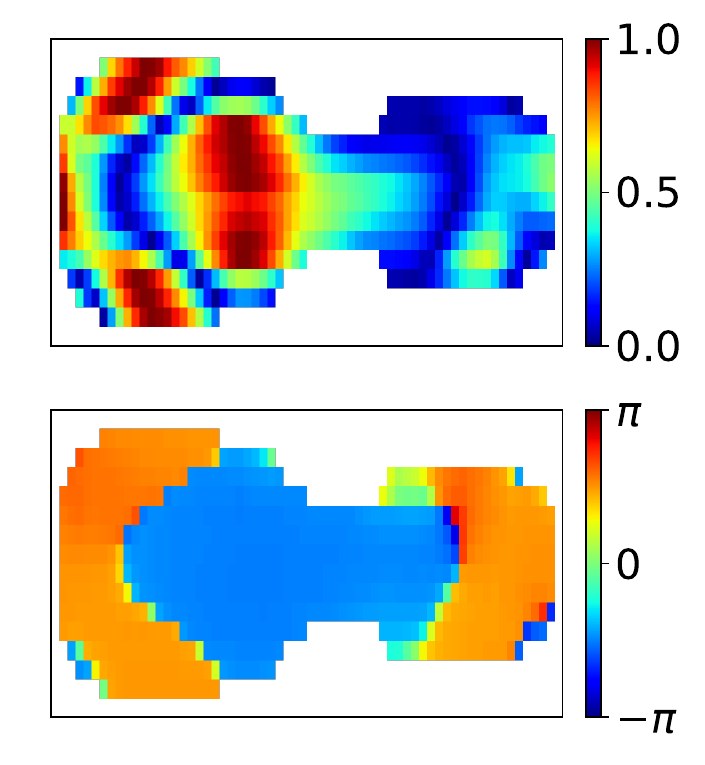}\label{fig:tl-53-khcnn}
    }
     \subfloat[\parbox{2cm}{\centering C-ESM\\ \tiny{NMSE: -0.55} \\ \tiny{NCC: 68.51\%}}]{
        \includegraphics[scale=0.28]{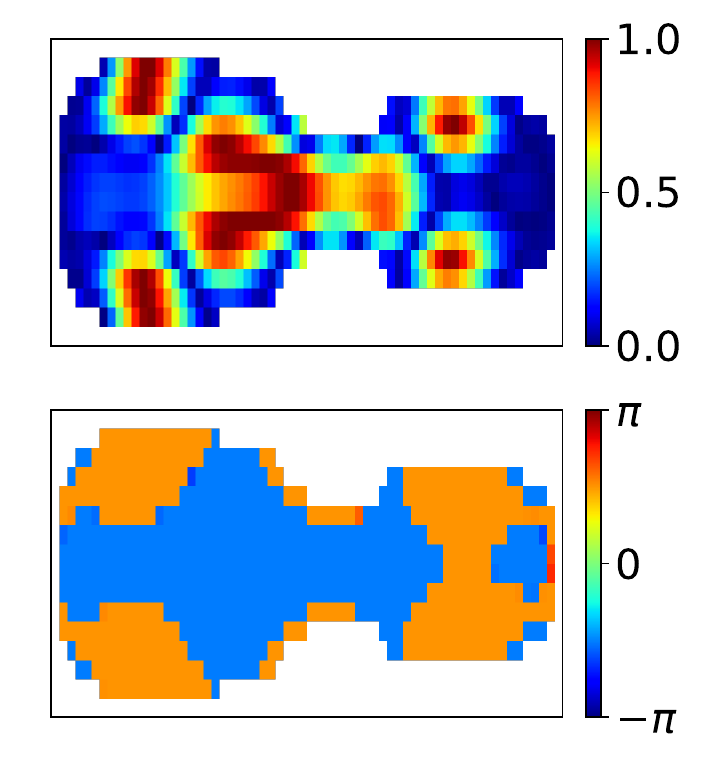}
    }
    \subfloat[\parbox{2cm}{\centering Fine-tuned (random init.) \\ \tiny{NMSE: -1.32} \\ \tiny{NCC: 53.88\%}}]{
        \includegraphics[scale=0.28]{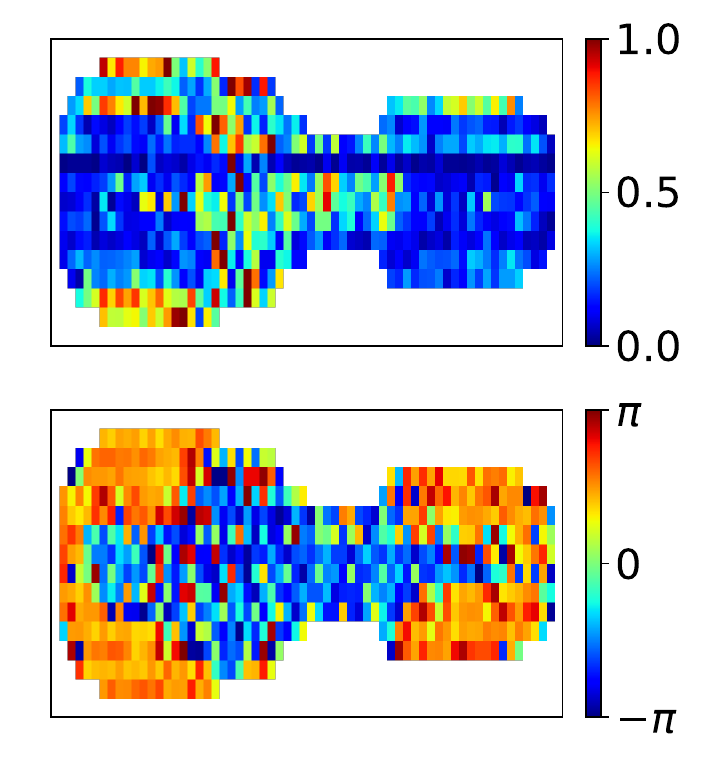}
    }

    \vspace{-0.3cm}
    \makebox[0pt][r]{%
        \rotatebox{90}{\hspace{1cm}\small Mode $26$, \SI{1255}{\hertz}}%
    }%
    \subfloat[Ground truth]{
        \includegraphics[scale=0.28]{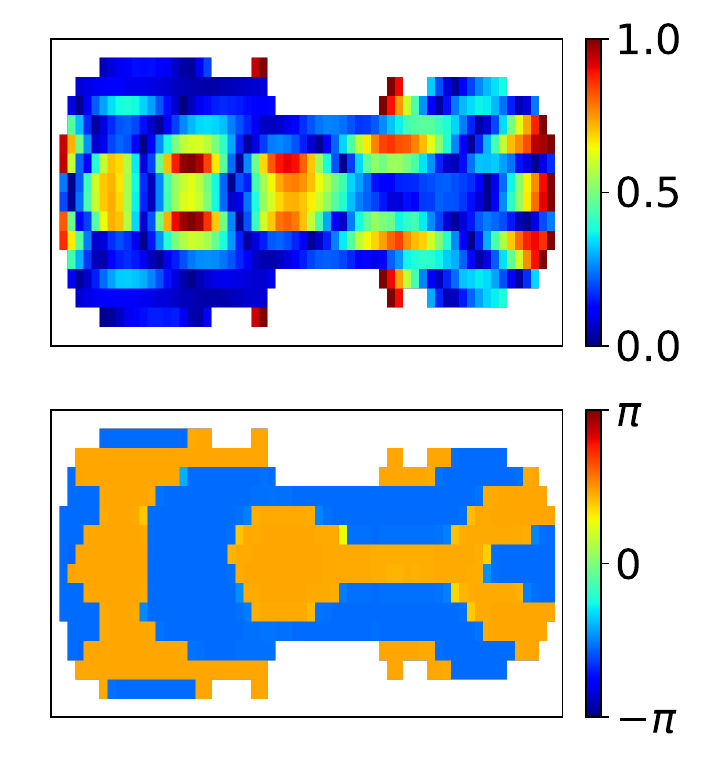}
    }
    \subfloat[\parbox{2cm}{\centering Fine-tuned \\ \tiny{NMSE: \textbf{-0.04}} \\ \tiny{NCC: \textbf{60.10\%}}}]{
        \includegraphics[scale=0.28]{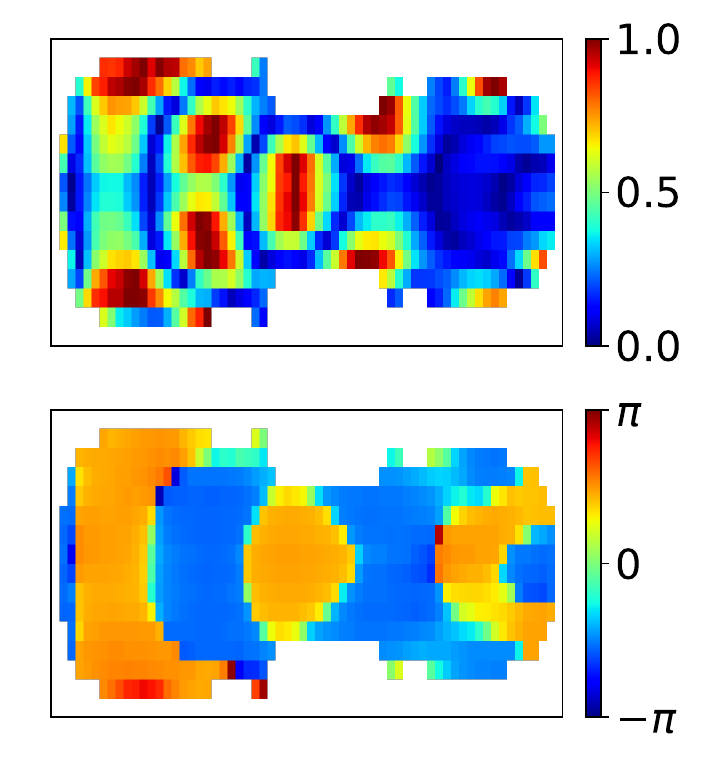}
    }
    \subfloat[\parbox{2cm}{\centering Pre-trained\\ \tiny{NMSE: 1.00} \\ \tiny{NCC: 52.39\%}}]{
        \includegraphics[scale=0.28]{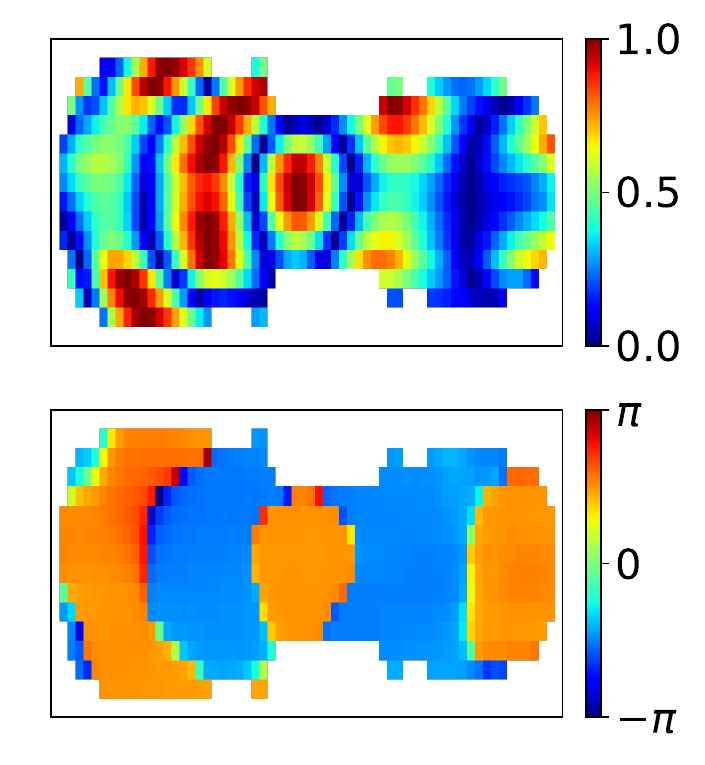}\label{fig:tl-53-khcnn}
    }
     \subfloat[\parbox{2cm}{\centering C-ESM \\ \tiny{NMSE: 2.80} \\ \tiny{NCC: 52.20\%}}]{
        \includegraphics[scale=0.28]{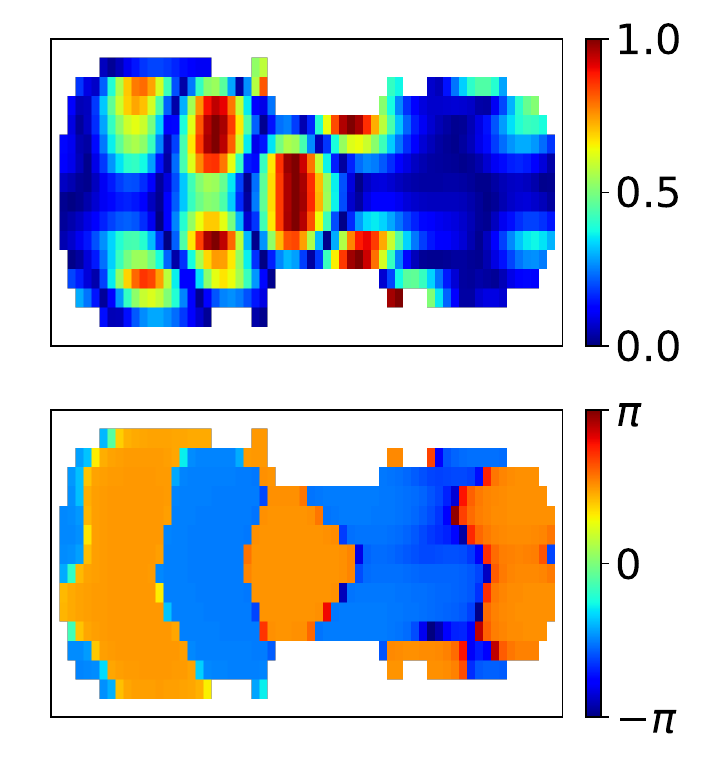}
    }
    \subfloat[\parbox{2cm}{\centering Fine-tuned (random init.)\\ \tiny{NMSE: -0.52} \\ \tiny{NCC: 47.38\%}}]{
        \includegraphics[scale=0.28]{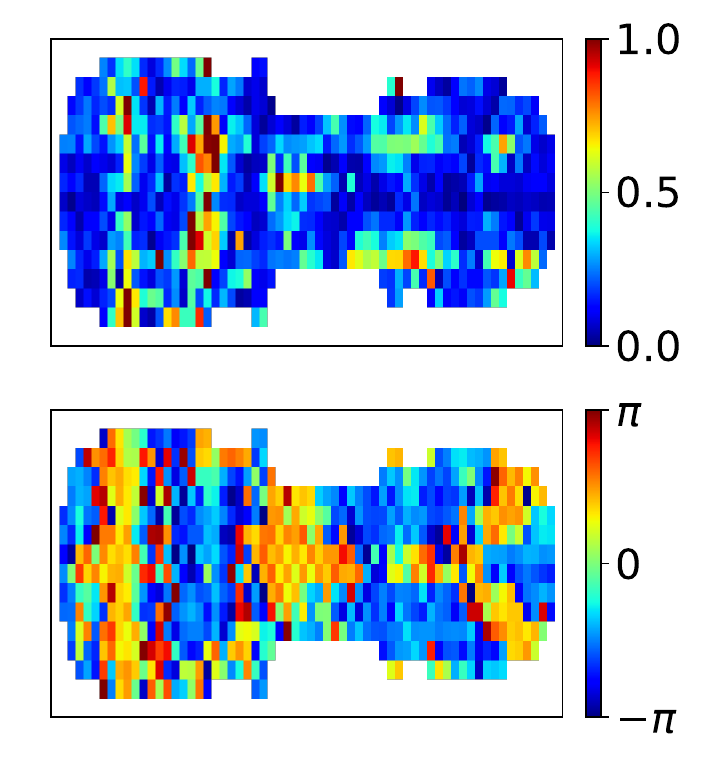}
    }
    \caption{Two examples of $\textbf{V}_\mathcal{S}$ for violin top plate dataset.}
    \label{fig:tl-sample}
\end{figure*}

\section{Proposed method}
We present a physics-informed transfer learning framework for a data-driven NAH approach. The framework comprises two stages: first, a supervised pre-training of a CV-CNN using a large dataset; and second, a physics-informed fine-tuning of the model using only a single data sample.

A U-Net-based complex-valued convolutional neural network (CV-CNN), denoted by $\Lambda(\boldsymbol{\theta})$ with $\theta$ representing the trainable parameters, maps the hologram pressure  $\mathbf{p}_\mathcal{H}$ to the  source velocity $\mathbf{v}_\mathcal{S}$, expressed as
\begin{equation}
        \hat{\mathbf{v}}_{\mathcal{S}}  = \Lambda(\boldsymbol{\theta})  {\mathbf{p}}_{\mathcal{H}}.
\end{equation}

The CV-CNN architecture is based on the design outlined in \cite{luancomplex}, with a modified U-Net \cite{ronneberger2015u} serving as the backbone. A key difference from the original model in \cite{luancomplex} is the use of a single decoder in this approach, which is dedicated to predicting source velocity. This architecture is maintained consistently throughout both the pre-training and fine-tuning stages. The network employs the Cardioid complex activation function, and its detailed structure is illustrated in Fig.~\ref{fig:TL-PINN framework}.

The core difference between the pre-training and fine-tuning lies in the loss function. During pre-training, the model is trained on a large dataset using a purely data-driven mean square error (MSE) loss, defined as 
\begin{equation}
    \mathcal{L}_{pre} = \frac{1}{N} \| {\mathbf{v}}_{\mathcal{S}} - \hat{\mathbf{v}}_{\mathcal{S}} \|_2^2. 
\label{eq:TL-pre-loss}
\end{equation}
Unlike CV-KHCNN \cite{luancomplex}, this stage does not incorporate any physics constraints, thereby omitting the computationally expensive backpropagation through the KH integral and significantly accelerating the training process. Note that due to the data-driven nature of the pre-training stage, it is necessary to have access to the ground-truth velocity of the sound source ${\mathbf{v}}_{\mathcal{S}}$.

Instead, the physics constraints are introduced during the fine-tuning stage. This stage can be regarded as a transfer learning process, where the trainable parameters are initialized with those from the pre-trained model, and the training is conducted on a specific data sample that lies outside the distribution of the pre-training dataset.
Since the pre-trained model operates on normalized data, the network output $\hat{\mathbf{v}}_{\mathcal{S}}$ is rescaled to its original physical scale using a trainable scaling factor $C$.

The fine-tuning stage is a self-supervised, physics-informed procedure that does not require knowledge of the ground-truth velocity of the sound source, ${\mathbf{v}}_{\mathcal{S}}$. In this stage, the loss function is defined as the mean absolute error (MAE) between the measured hologram pressure and the one computed via the KH integral \eqref{eq:kirchhoffhelmholtz} from the predicted source velocity:
\begin{equation}
    \mathcal{L}_{tune} = \frac{1}{M} \| {\mathbf{p}}_{\mathcal{H}} - \hat{\mathbf{p}}_{\mathcal{H}} \|_1  .
\label{eq:TL-fine-loss}
\end{equation}
Empirically, we find that adopting MAE results in faster and more stable convergence than MSE during fine-tuning. 
This choice is also motivated by the strategy employed in \cite{luan2025pinnsfd}.
It is worth mentioning that reconstructing the sound source directly at the actual source plane (as in \eqref{eq:TL-fine-loss}) typically leads to issues of non-uniqueness and singularity \cite{koopmann1989method}. These challenges can be alleviated by reconstructing an equivalent source located internally, behind the actual source, as done in the ESM \cite{koopmann1989method}.
However, to address these issues, we preserve the knowledge gained during pre-training by initializing the fine-tuning model with the pre-trained parameters and applying a small learning rate, ensuring that the network weights remain close to their initial values.
This transfer learning strategy enables rapid adaptation to the new dataset.

The pre-trained network is optimized using the Adam optimizer with an initial learning rate of $0.01$. The learning rate is reduced by a factor of $0.1$ after a plateau of $5$ epochs without improvement, down to a minimum of $0.0009$. To prevent overfitting, we apply early stopping, terminating training if no improvement in the validation loss is observed for $20$ consecutive epochs.
During the fine-tuning stage, two separate Adam optimizers are used: one for updating the neural network parameters with a fixed learning rate of $1 \times 10^{-3}$, and another for adjusting the scaling factor with a learning rate of $1 \times 10^{-5}$. The model undergoes fine-tuning for a total of $2000$ epochs.
\textit{Pytorch} \cite{paszke2019pytorch} is used for the implementation and the library \textit{complexPyTorch} \cite{matthes2021learning} is used to implement CVNNs in this study. 

%% file: sections/validation.tex
\begin{table}[tb!]
\begin{center}
\caption{The mean value of $\operatorname{NMSE}$ and $\operatorname{NCC}$ of $\hat{\textbf{V}}_\mathcal{S}$ and runtime for different models.  }
\label{table:TL-PINN}
\begin{tabular}{l c c c} 
\hline
\textbf{} & $\operatorname{NMSE}$ &  $\operatorname{NCC}$ & runtime \\
\hline
pre-trained & -0.33 & 54.52\%  & \SI{4.09}{\hour} \\
fine-tuned &  {-1.76} & 60.66\% & \SI{1.28}{\minute} per sample\\
C-ESM &  -1.13& {63.20\%} & -\\
\hline
\end{tabular}
\end{center}
\vspace{-3mm}
\end{table}
\section{Validation}
\subsection{Implementation}
The NAH configuration is as follows:
the surface plane is located $z_\mathcal{H} = \SI{3.12}{\centi\meter}$ away from the hologram plane.
The grid on $\mathcal{H}$ contains $M = 8 \times 8$ measurement points, while the source field $\mathcal{S}$ is discretized with $N= 16 \times 64$ points, representing an up-sampling from the measurement plane.

The proposed approach is validated in a scenario where a large dataset is available and we aim to transfer the learned knowledge to a different, limited-data setting. Specifically, the model is first pre-trained on a simulated dataset of rectangular plates with varying boundary conditions, which is an abundant and easily generated dataset. It is then fine-tuned using data from a violin top plate, which has an arbitrary shape and is difficult to acquire. 
A simulated NAH dataset, as described in \cite{olivieri2021pinn}, is used and includes objects such as rectangular plates and violin top plates. 
The eigenfrequency for the dataset generation is limited in $[0, 2000]$ $\si{\hertz}$. 
The pre-trained rectangular plate dataset is divided into training, validation, and test sets in an $8:1:1$ ratio, while separate fine-tuning is performed for all modes on $10$ different violin top plates ($442$ samples in total). The dataset pre-processing follows the same approach as described in \cite{olivieri2021pinn}.
The GPU utilized is an NVIDIA GeForce RTX 4080 with 16GB VRAM. 
Furthermore, the conventional C-ESM method \cite{fernandez2017sparse} is also implemented  on the test dataset of violin top plates, using the MATLAB toolbox CVX \cite{grant2014cvx} to serve as a benchmark. Note that C-ESM is not a training-based deep learning approach. Five regularization parameters are applied, evenly spaced within the range of $[0.001, 0.1]$. The final result is selected based on the best reconstruction of the hologram pressure field with the minimum MAE loss. 
\subsection{Results}
 The performance of the proposed approach is assessed by two metrics: the Normalized Mean Square Error ($\operatorname{NMSE}$) and the Normalized Cross Correlation ($\operatorname{NCC}$). 
 They are expressed by
\begin{equation}
    \operatorname{NMSE}(\hat{\mathbf{x}},\mathbf{x}) = 10\mathrm{log}_{10}\left ( \frac{\mathbf{e}^H \cdot \mathbf{e}}{\mathbf{x}^H \cdot \mathbf{x}} \right ), 
    \operatorname{NCC}(\hat{\mathbf{x}},\mathbf{x}) = \frac{\hat{\mathbf{x}}^H \cdot \mathbf{x}  }{\|\hat{\mathbf{x}} \|_2 \cdot \| \mathbf{x}\|_2},
\end{equation}
where $\mathbf{x}$ are the ground truth data, $\hat{\mathbf{x}}$ are the predicted data, and $\mathbf{e} = \hat{\mathbf{x}} - \mathbf{x}$ denote the error. Additionally, the metrics are computed with a column-vector representation of the data, and $\operatorname{NCC}$ reaches 1 when the two quantities positively correlate perfectly.
Note that both the metrics are for complex numbers and $^H$ is the Hermitian transpose operator.
Since the violin plate has irregular shape, binary mask is adopted to select the points of the mesh grid belonging to the target surface when evaluating the surface plane of the violin top plates, as done in \cite{olivieri2021pinn}.

The $\operatorname{NMSE}$ and $\operatorname{NCC}$ results for the violin top plates using the pre-trained model, fine-tuned model, and C-ESM are presented in Table~\ref{table:TL-PINN}. 
Note that the pre-trained model is evaluated directly on the violin top plate dataset, which is identical to the data used for fine-tuning, rather than on the original rectangular plate training dataset. All data are normalized to facilitate comparison with the pre-trained model, which outputs normalized results by design.
The corresponding runtimes for training the pre-trained and fine-tuned models are also included in Table~\ref{table:TL-PINN}.
The results indicate that fine-tuning improves the model’s accuracy compared to the pre-trained model, demonstrating the effectiveness of the fine-tuning procedure. However, when comparing the fine-tuned model with the C-ESM, the accuracy remains competitive rather than showing a clear improvement.
On the other hand, the fine-tuning procedure requires significantly less time compared to pre-training, indicating that the inference time of the fine-tuned model may be acceptable and promising for real-world applications.

Two representative examples are presented in Fig.~\ref{fig:tl-sample}, showing the ground truth along with the reconstructions from the fine-tuned model, pre-trained model, C-ESM, and a fine-tuned model with random initialization. The top sample illustrates a case where the pre-trained model already yields a good reconstruction (and better than C-ESM), which is further improved through fine-tuning. The bottom sample depicts a scenario where the pre-trained model and C-ESM produce comparable results, but fine-tuning still enhances the accuracy. Notably, the phase in the upper bout is adjusted to more closely match the ground truth. The result from the randomly initialized fine-tuned model demonstrates that directly optimizing the loss function in \eqref{eq:TL-fine-loss} on the source plane without pre-training can lead to issues of nonuniqueness and singularity. 
\begin{figure}[t!]
\vspace{-0.8cm}
    \centering
    \subfloat[]{
        \includegraphics[width=0.48\linewidth]{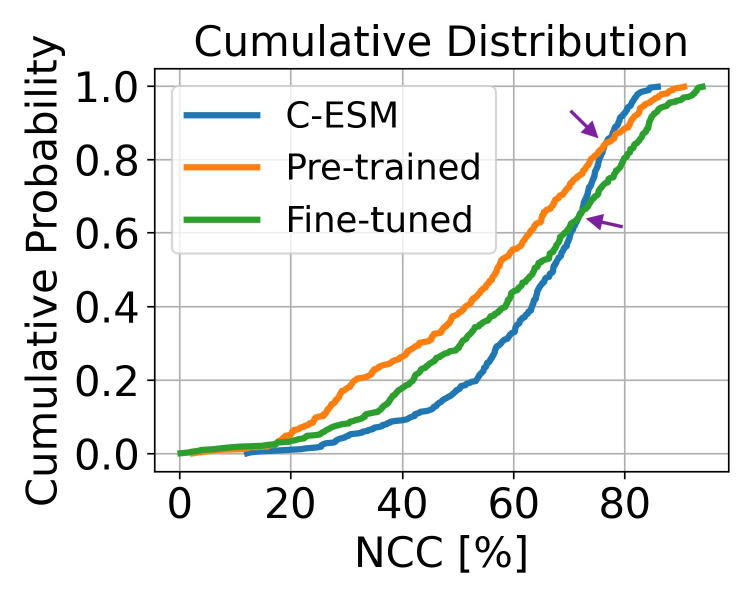}
    }
    \subfloat[]{
        \includegraphics[width=0.48\linewidth]{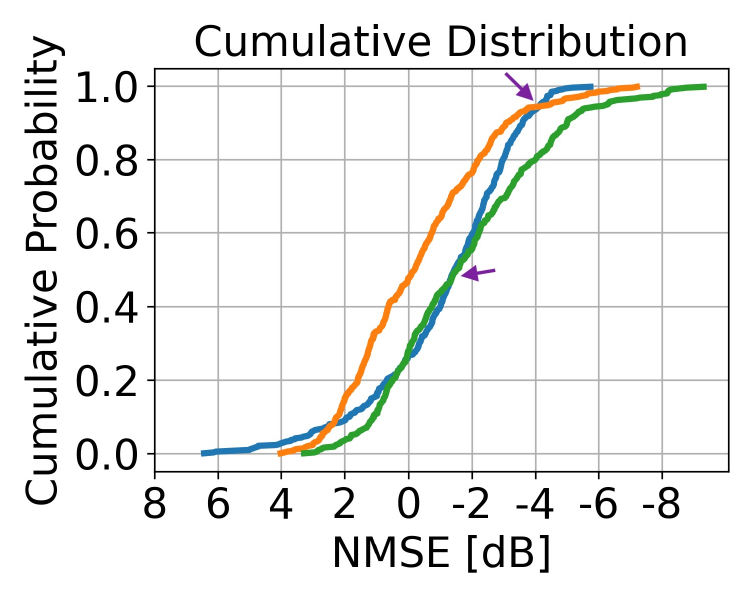}
    }
    \caption{Cumulative distribution for $\operatorname{NCC}$ and $\operatorname{NMSE}$.}
    \label{fig:cumu}
\end{figure}
To gain deeper insight into the distribution of $\operatorname{NMSE}$ and $\operatorname{NCC}$, we also evaluate their cumulative distribution functions $P(X \leq x)$, which is the probability that a random variable $X$ takes a value less than or equal to $x$. The cumulative distribution for $\operatorname{NCC}$ and $\operatorname{NMSE}$ from C-ESM, pre-trained and fine-tuned are shown in Fig.~\ref{fig:cumu}. Note that to align the trends of $\operatorname{NCC}$ and $\operatorname{NMSE}$, the $\operatorname{NMSE}$ is accumulated in descending order, from higher to lower values, ensuring that the accumulation starts from the poorer reconstruction cases.
At a given cumulative probability, a higher $\operatorname{NCC}$ or a lower $\operatorname{NMSE}$ indicates better reconstruction performance.
We observe that the fine-tuned and C-ESM curves intersect in regions corresponding to high $\operatorname{NCC}$ and low $\operatorname{NMSE}$, as highlighted by the arrows in the plots.
Below the intersection point, C-ESM outperforms (with higher $\operatorname{NCC}$ and lower $\operatorname{NMSE}$) the fine-tuned model at the same cumulative probability, whereas beyond the intersection, the fine-tuned model achieves better performance. 
That said, our primary focus may be on the successful modes. In other words, modes with reconstruction quality below a certain threshold can be considered failures.  In these cases, it becomes irrelevant to claim that a 30\% reconstruction is superior to a 20\% one in terms of $\operatorname{NCC}$, as both are insufficient for meaningful interpretation.
Thus, it is more meaningful to concentrate on the successful modes, where the fine-tuned approach consistently demonstrates higher accuracy above the intersection. Additionally, another intersection is observed between the pre-trained model and C-ESM (located to the right of the previous one), indicating that even without fine-tuning, the pre-trained model, which was trained on the rectangular plate dataset, can outperform C-ESM in terms of accuracy for the successful modes.

\begin{figure}[t!]
\vspace{-0.8cm}
    \centering
    \subfloat[]{
        \includegraphics[width=0.48\linewidth]{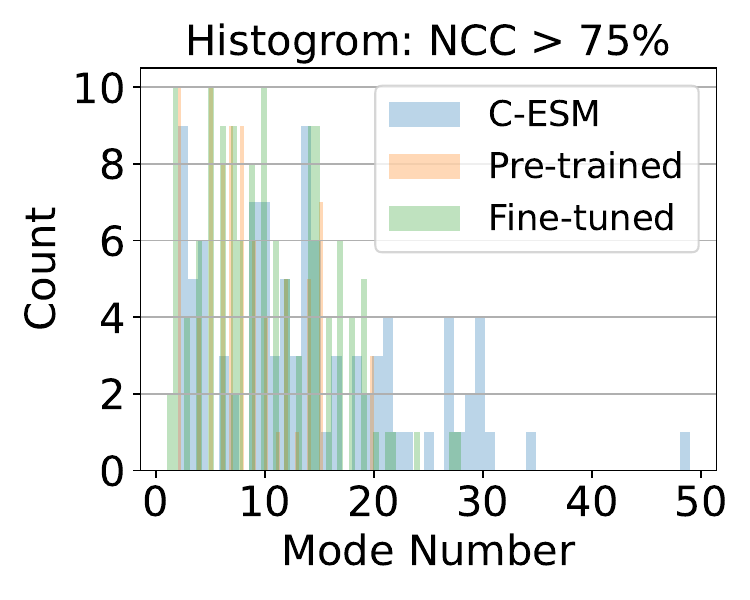}
    }
    \subfloat[]{
        \includegraphics[width=0.48\linewidth]{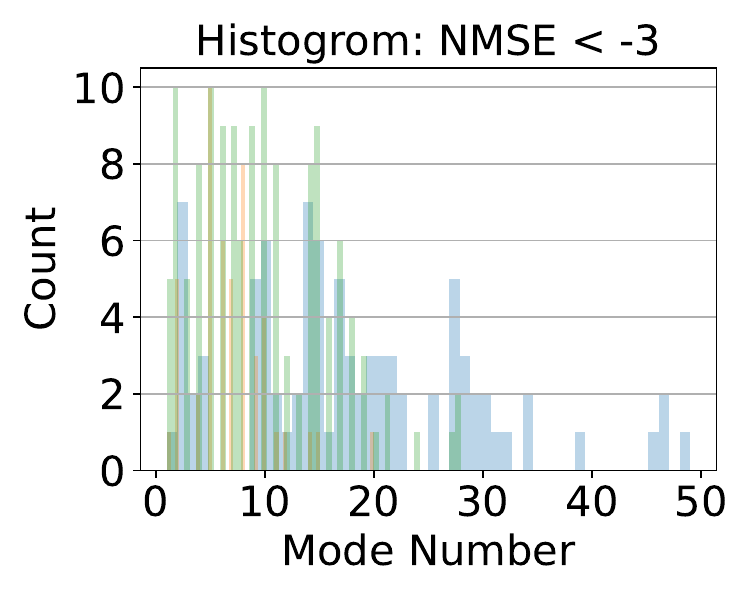}
    }
    \caption{Histogram of successful modes $\operatorname{NCC}$ and $\operatorname{NMSE}$.}
    \label{fig:hist}
\end{figure}

We also provide histograms of $\operatorname{NCC}$ and $\operatorname{NMSE}$ for the successful modes ($\operatorname{NCC} > 75\%$  and $\operatorname{NMSE} < -3$), categorized by mode number, as shown in Fig.~\ref{fig:hist}. The results reveal that most successful reconstructions correspond to lower mode numbers. Additionally, both the pre-trained and fine-tuned models show a higher concentration of low-mode success and fewer high-mode successes compared to C-ESM.
This indicates that the fine-tuned model performs better on lower modes, which we suspect is due to significant discrepancies in high frequency mode shapes between the violin top plate and the rectangular plate. 

%% file: sections/conclusion.tex
\section{Conclusion}\label{sec:conclusion}

In this work, we propose a transfer learning framework for sound source reconstruction in NAH, which integrates physics-informed procedures to adapt a pre-trained model from one type of sound source to another. The framework consists of two stages: (1) supervised pre-training of a CV-CNN on a large dataset, and (2) physics-informed fine-tuning on a single data sample using the KH integral. 
This approach mitigates the issues of nonuniqueness and singularity associated with direct source field reconstruction on the source plane by leveraging the representational power of deep neural networks to capture the complex relationship between hologram measurements and source distributions. A model pre-trained on a more accessible dataset can successfully transfer its learned knowledge, enhancing reconstruction performance on more complex or limited datasets.

The framework is validated by transferring a pre-trained model from a rectangular plate dataset to a violin top plate dataset, where it demonstrates improved reconstruction accuracy over the pre-trained model and achieves competitive performance relative to the C-ESM. Additionally, for successful modes, the fine-tuned model outperforms both the pre-trained model and C-ESM. It also demonstrates superior performance specifically on low-frequency mode shapes.

Although we validate the approach on a particularly challenging scenario, given the expected differences between a rectangular plate and a violin top plate, the results still demonstrate its effectiveness. This physics-informed transfer learning approach shows strong potential and could attract greater attention as it continues to be developed for real-world applications.